\documentclass[structabstract]{aa}
\usepackage{natbib}
\bibpunct{(}{)}{;}{a}{}{,}
\usepackage{graphicx}
\usepackage{txfonts}

\defcitealias{kis11b}{Paper~I}

\begin{document}
\unitlength1cm

\title{VLTI/AMBER observations of the Seyfert nucleus of NGC~3783
\thanks{Based on observations made with ESO telescopes at Paranal Observatory under programme IDs 083.B-0212(A) and 087.B-0578(A).}   }

\author{ G.~Weigelt \inst{1} 
\and K.-H.~Hofmann \inst{1}  
\and M.~Kishimoto  \inst{1}         
\and S.~H\"onig \inst{2} 
\and D.~Schertl  \inst{1}
\and  A.~Marconi \inst{3,4}    
\and  F.~Millour \inst{5}    
\and R.~Petrov \inst{5}
\and D.~Fraix-Burnet \inst{6}
\and F.~Malbet \inst{6}     
\and K.~Tristram \inst{1}
\and M.~Vannier  \inst{5}          }

\offprints{G.~Weigelt\\ email: \texttt{weigelt@mpifr.de}}
\institute{Max-Planck-Institut f\"ur Radioastronomie, Auf dem H\"ugel 69, D-53121 Bonn, Germany
\and UCSB Department of Physics, Broida Hall 2015H, Santa Barbara, CA, USA
\and Dipartimento di Fisica e Astronomia, Universit\'a di Firenze, Largo Enrico Fermi 2, 510125, Firenze, Italy
\and INAF - Osservatorio Astrofisico di Arcetri, Largo Fermi 5, 50125, Firenze, Italy
\and Laboratoire Lagrange, UMR7293, Universit\'e de Nice Sophia-Antipolis, CNRS, Observatoire de la C\^ote d'Azur, 06300 Nice, France
\and Universit\'e Joseph Fourier (UJF) - Grenoble 1 / CNRS-INSU, Institut de Plan\'etologie et d'Astrophysique de Grenoble (IPAG) UMR 5274, Grenoble, F-38041, France           }

\authorrunning{G.~Weigelt et al.\ }
\titlerunning{VLTI/AMBER observations of the Seyfert AGN NGC~3783}

\date{Received ; accepted }

\abstract
{The putative tori surrounding the accretion disks of active galactic nuclei (AGNs) play a fundamental role in the unification scheme of AGNs.  Infrared long-baseline interferometry   allows us to study the inner dust distribution in AGNs with unprecedented spatial resolution over a wide infrared wavelength range. }
{Near- and mid-infrared interferometry is used to  investigate the milli-arcsecond-scale dust distribution in the type 1.5 Seyfert nucleus of NGC~3783.    }
{We observed NGC~3783 with the VLTI/AMBER instrument in the $K$-band and compared our  observations with models. }
{From the $K$-band observations, we derive a ring-fit torus radius of $0.74 \pm 0.23$~mas or $0.16 \pm 0.05$~pc. 
We compare this size with infrared interferometric observations of  other AGNs and UV/optical-infrared reverberation measurements.  For the interpretation of our observations, we simultaneously model our near- and   mid-infrared visibilities and the SED  with a temperature/density-gradient model including an additional inner hot 1400~K ring component. 
 }    {}            
   
\keywords{galaxies: active -- galaxies: Seyfert -- infrared: galaxies -- techniques: interferometric -- galaxies: individual: NGC~3783}

\maketitle

\begin{figure}                      
  \centering
  \par\vspace{3mm}  
 \hspace*{-0mm}
    \includegraphics[height=66mm,angle=270]{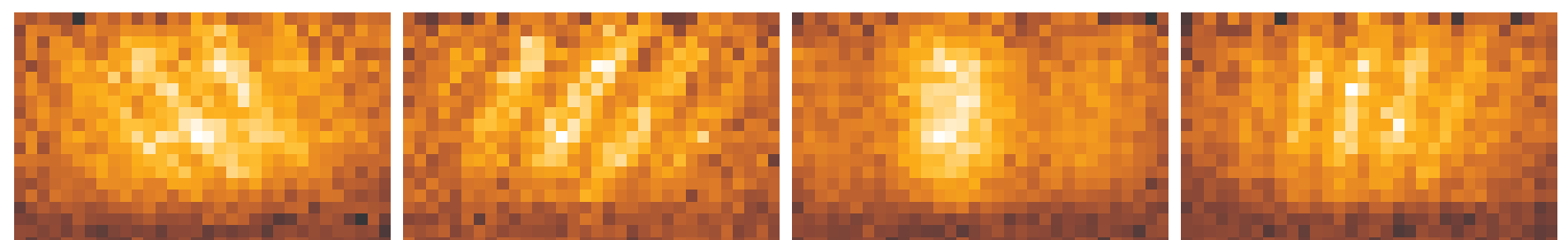} \\   
      \par\vspace{3pt}    
      \hspace*{-3mm}
    \includegraphics[height=42mm,angle=0]{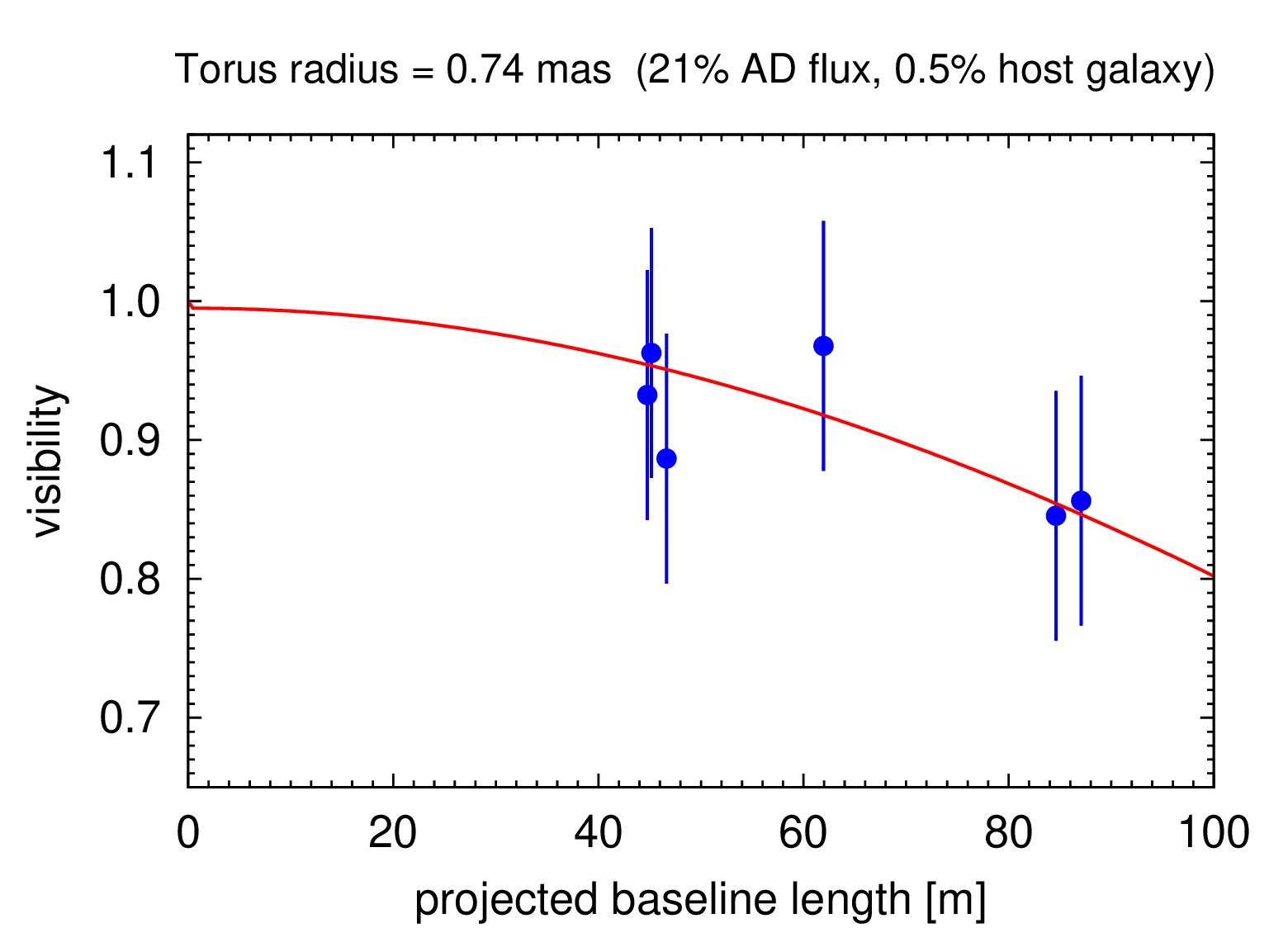}
     \par\vspace{-2pt}    
     \hspace*{-5mm}
          \includegraphics[height=61mm,angle=270]{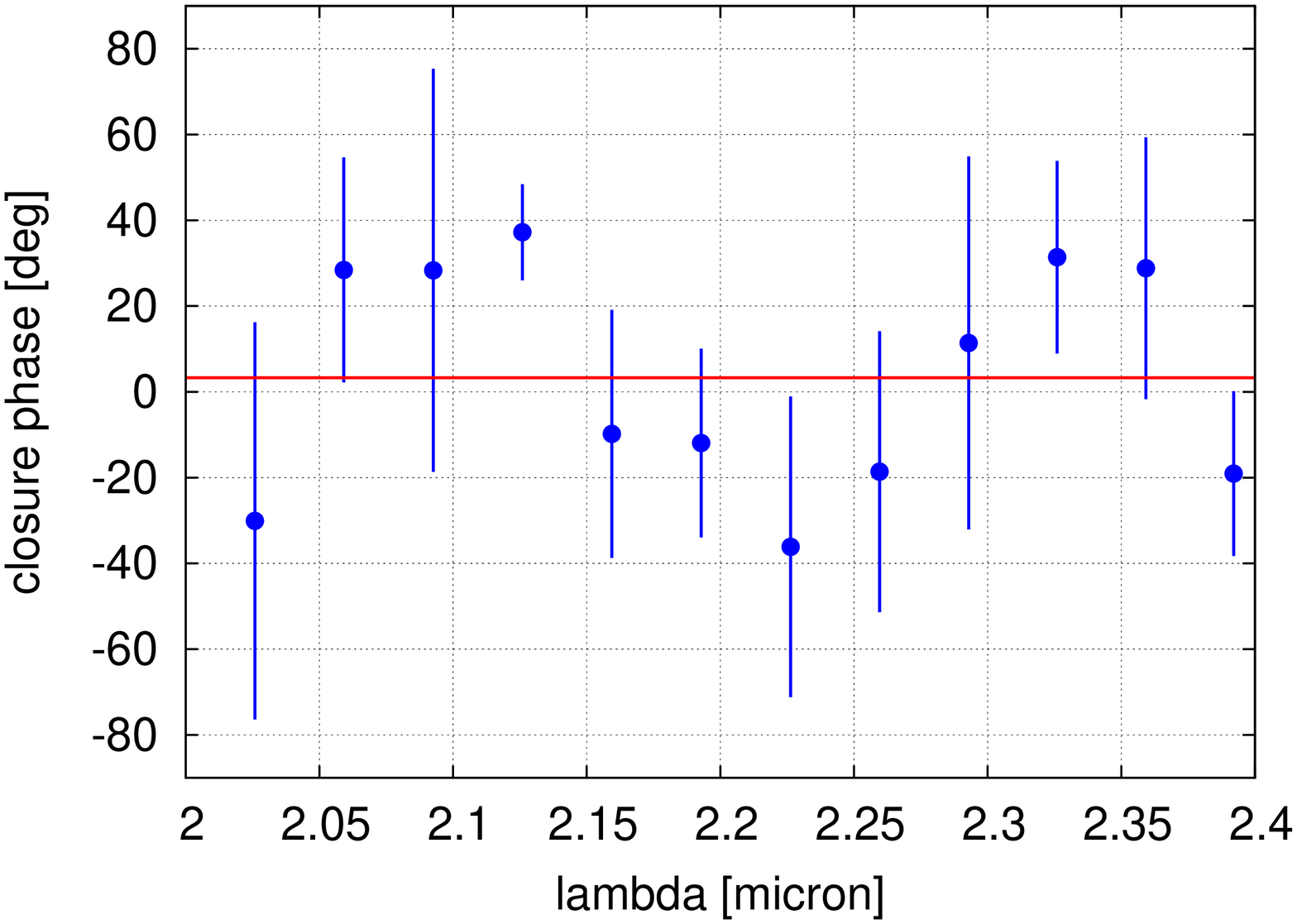}
          \par\vspace{0pt}
      \caption[]{\small            
{\it Top:} Examples of LR $K$-band AMBER interferograms of NGC~3783 ($\sim$2.0--2.4\,$\mu$m from bottom to top), which illustrate the noise problem. From left to right, the first and second interferogram are recorded with UT2--UT3  (46.6~m projected baseline) with  DIT = 800~ms and 400~ms,  respectively, the third one with UT3--UT4 (62.5~m, 800~ms), and the last one with UT2--UT4 (89.4~m, 800~ms). 
{\it Middle:} Calibrated visibilities of NGC~3783 and geometric ring-model fit (red). We derive a ring-fit torus radius of $0.74 \pm 0.23$~mas or $0.16 \pm 0.05$~pc. 
{\it Bottom:} Closure phases plotted versus wavelength.
 }
\label{fig:obs}
\end{figure}

\section{Introduction}\label{sec1}
The putative tori surrounding the accretion disks of active galactic nuclei (AGNs) play a fundamental role in the unification scheme of AGNs  \citep{ant93}. These tori most likely provide the material reservoir that feeds the accretion disk (e.g., \citealt{kro88}). Infrared interferometric observations can resolve  these mas-scale  tori  in the near-infrared (NIR) (e.g., \citealt{swa03,wit04,kis09a,pot10,kis11a})  and mid-infrared (MIR) (e.g., \citealt{jaf04,mei07,tri07,tri09,bec08,kis09b,rab09,bur09,bur10,tri11}, and \citealt[][Paper~I]{kis11b}). In this paper, we present the first NIR interferometric observation of an AGN  with the AMBER/VLTI instrument.  

\begin{table*}                  
\caption{Observation log of our AMBER LR observations of NGC~3783  and its calibrator CD-37 7391. The data in the first 11 lines were observed in the two-telescope mode  and the data in the last 4 lines in the three-telescope mode. The table lists the names, times of observations, projected baseline lengths, position angles PA, detector integration times DIT, seeing, number of interferograms, and derived target visibilities (errors $\pm0.09$).}

{\tiny    \begin{tabular}{lllllrrrr} \hline \hline
Name            & date           & Time of                & Telescopes/                     & PA                     & DIT     & Seeing     & Number     &Target \\
                      &                  & observation          & proj. baseline                  & (\degr)               &(ms)     & (arcsec)   & of frames   &visibility\\
                      &                  & (UTC)                   & lengths (m)                     &                           &           &                 &                            \\ \hline
NGC 3783     & 09/04/14   & 01:19 - 01:31       & UT2-3/46.6                      & 28.8                   & 800     & 0.93         & 11x70        &0.89\\
CD-37 7391   & 09/04/14   & 01:45 - 01:55      & UT2-3                              &                            & 800     & 0.84         & 10x70       &\\
NGC 3783     & 09/04/14   & 03:16 - 03:29      & UT2-3/45.2                       & 44.8                    & 800     & 0.82         & 11x70       &0.96\\
NGC 3783     & 09/04/14   & 03:32 - 03:41       & UT2-3/44.8                      & 46.3                    & 400     & 0.79         & 10x70       &0.93\\
CD-37 7391   & 09/04/14   & 04:31 - 04:34      & UT2-3                               &                            & 400     & 0.78         & 4x120       &\\ \hline
NGC 3783     & 09/04/14    & 03:52 - 04:03     & UT3-4/62.0                       & 120.9                  & 800     & 0.68         & 10x70       &0.97\\
CD-37 7391   & 09/04/14   & 02:11 - 02:16      & UT3-4                               &                            & 800     & 1.45         & 5x70         &\\ \hline
CD-37 7391   & 09/04/14   & 02:00 - 02:05      & UT2-4                               &                           & 800      & 0.99         & 5x120       &\\
NGC 3783     & 09/04/14   & 03:44 - 03:49      & UT2-4/87.1                       & 90.0                    & 800     & 0.72         & 5x70          &0.86\\
NGC 3783     & 09/04/14    & 04:12 - 04:20     & UT2-4/84.6                       & 94.7                    & 800     & 0.81         & 8x70          &0.85\\
CD-37 7391   & 09/04/14   & 04:41 - 04:45     & UT2-4                                &                            & 800     & 0.87         & 3x120        &\\  \hline
NGC 3783     & 11/05/18    & 02:39 - 02:44     & UT1-2-4/51.2/113.8/80.3   & 38.8/77.3/100.7  & 400     & 0.63         & 7x120        &\\ 
NGC 3783     & 11/05/18    & 02:50 - 02:55     & UT1-2-4/50.6/111.2/78.7    & 39.8/78.9/102.8  & 400     & 0.68         & 7x120        &\\ 
CD-37 7391   & 11/05/18    & 02:00 - 02:06     & UT1-2-4/                            &                            & 400     & 0.75         & 7x120        &\\ 
CD-37 7391   & 11/05/18    & 02:09 - 02:17     & UT1-2-4/                            &                            & 400     & 0.63         & 7x120        &\\ \hline 
 \end{tabular}       }
\label{tab:obslog}
\end{table*}

\section{Observations and data reduction} \label{sec2}
We observed the Seyfert 1.5 AGN NGC~3783 in 2009 and 2011 (IDs 083.B-0212 and 087.B-0578) with  the ESO VLTI and the AMBER instrument  \citep{pet07}.  For these observations in the $K$-band (see Table~\ref{tab:obslog}), the AMBER low spectral resolution mode (LR) was employed. Figure~\ref{fig:obs} (top) presents examples of target interferograms  to illustrate the noise problem  ($K \sim$ 10.1).  Long detector integration times  (DIT) of 400 and 800~ms were chosen to be able to recognize the faint fringes during data recording and correct  drifts of the optical path differences  (OPDs) between the telescope beams.  The interferograms shown in Fig.~\ref{fig:obs}  are two-telescope interferograms, not AMBER-standard three-telescope interferograms. In 2009 (see Table~\ref{tab:obslog}), we recorded two-telescope interferograms since it is easier to correct OPD drifts in two-telescope  than in three-telescope interferograms. In 2011, we recorded three-telescope interferograms, which provide closure phases. For data reduction of the two-telescope interferograms, we used our own  software  (developed by one of us, KHH), which is able to reduce the non-standard two-telescope interferograms. It is based on the same P2VM algorithm \citep{tat07,che09} as the AMBER   \textit{amdlib}\footnote{The AMBER - reduction package \textit{amdlib} is available at: http://www.jmmc.fr/data\_processing\_amber.htm}  software. To reduce the effect of the instantaneous  OPDs on the visibility, we applied a preprocessing method that equalizes the OPD histograms of the target and calibrator interferograms \citep{kre12}. Figure~\ref{fig:obs}  and Table~\ref{tab:obslog} show the derived  visibilities, which are wavelength averages over the wavelength range of 2.0--2.40~$\mu$m. For data reduction of the 2011 three-telescope data, we used the standard AMBER data reduction package \textit{amdlib} version 3.0. Figure~\ref{fig:obs} (bottom) shows the closure phases of NGC~3783 derived from the three-telescope 2011  interferograms. The average closure phase is $3.3 \pm 26\degr$. Closure phases are a measure of asymmetry. However, the large errors do not allow us to detect any asymmetry.  We also tried to derive calibrated visibilities from the 2011 data, but without success since the transfer function was unstable. 

\begin{table*}
\caption[]{NGC~3783 torus radius $R_{\rm torus}$, 2MASS fluxes of the nuclear core, and flux contributions of the host galaxy and the AD.      } 
{\tiny     \begin{tabular}{lccccccccccccccccccc}
\hline \hline
 $J$ flux   &$H$ flux  &$K$ flux &$J$        &$H$      &$K$       &host                       & AD                 &$R_{\rm torus}$  & $R_{\rm torus}$  & $R_{\rm \tau_K} ^c$\\ 
 (mJy)      & (mJy)     & (mJy)    & (mag.)  & (mag.)  & (mag.)  & fraction$^a$         & fraction$^b$  & (mas)                   & (pc)                   & (pc) \\ 
\hline
 18.8       & 34.2        & 61.8       & 12.3     & 11.2      & 10.1     & 0.005$\pm$0.002  & 0.21$\pm$0.07  & $0.74\pm0.23$  & $0.16\pm0.05$  &0.071$\pm$0.025  \\ 
\hline
\end{tabular}    }    \\
$^a$$K$-band flux contribution in the AMBER  FOV. $^b$AD flux contribution to the point-like core in the 2MASS  $K$-band image. 
$^c$Reverberation radius $R_{\rm \tau_K}$  \citep{gla92}.
\label{tab_2mass}  
\end{table*}

\section{Geometric model fits}    To interpret our $K$-band visibilities (Fig.~\ref{fig:obs} middle), we first fitted a geometric thin-ring model (i.e.,  ring width = outer radius minus inner radius = 0) to the visibilities and derived a ring-fit radius of $\sim$0.67~mas. This is only a very rough estimate of the torus size  since the observed visibilities may not only depend on the torus but also on the  underlying galaxy within the 60\,mas field-of-view (FOV) of AMBER and on the accretion disk (AD),  which is thought to remain unresolved (\citealt{kis07,kis09a,kis09b}). Therefore,  we have to estimate  the  flux contributions from the  host galaxy and the AD point source and take these contributions into account when fitting the visibilities. 

From the $K$-band image of NGC~3783 in the 2MASS catalog,  we estimated the flux contribution of the host galaxy within the 60~mas AMBER FOV to $0.5 \pm 0.2$\,\%. To obtain the NIR flux from the torus and the AD, we used two-dimensional fits to separate the point-like core component in the 2MASS $J$-, $H$-, and $K$-band images from the underlying host galaxy (see Table~\ref{tab_2mass}), following the same procedure as described by \citet{kis09a}.  Using the derived NIR core fluxes, we can estimate the flux contribution of the AD component in the $K$ band.  We assume here that the core component flux originates from the hot dust and from the  AD.  Therefore, we fitted a power-law spectrum for the AD  and a blackbody  for the dust emission, as described in \citet{kis09a}.  We also applied a small correction for Galactic reddening with $E_{B-V}$ = 0.119. By assigning an uncertainty of the NIR AD spectral index of 0.3, we also obtained the uncertainty of the $K$-band AD flux contribution.  The AD flux fraction in the $K$-band was estimated to be as small as $21 \pm 7$\%, which is similar to the values of several other AGNs reported by \citet{kis07,kis09a}.  If we now take into account  these estimated flux contributions of $\sim$21\%  from the unresolved AD and of $\sim$0.5\% from the  host galaxy, we can derive the visibilities of the torus itself and can fit the radius of the torus  (i.e., this radius is the only fit parameter; the AD contributes  just a constant of 0.21 to the total visibilities). We derive a torus radius  $R_{\rm torus}$ of $0.74 \pm 0.23$~mas  or $0.16\pm0.05$~pc   (thin-ring fit; see Fig.~\ref{fig:obs} middle, red curve).  
                                                                                    
\section{Interpretation and discussion}
\subsection{NIR interferometric and reverberation radii }

\begin{figure}      
\centering \includegraphics[width=7.8cm]{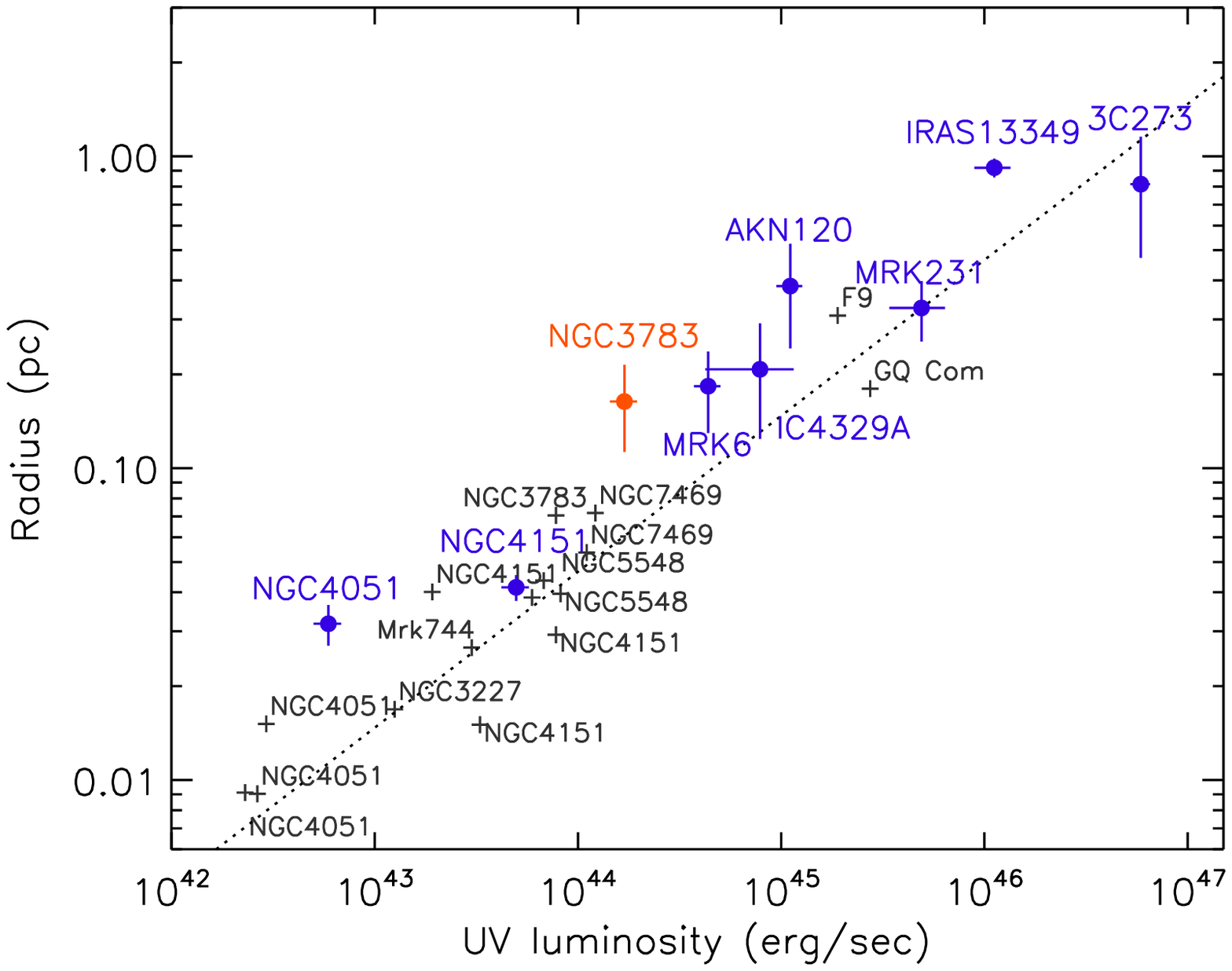}
\caption{$K$-band torus radii   of  NGC~3783 (red dot) and eight other AGNs  (blue; from \citealt{kis11a})   versus their UV luminosities.  The black symbols and the dotted line are the reverberation radii $R_{\rm \tau_K}$ \citep{gla92,sug06} and their fit curve, respectively.}
\label{fig_rev_ring}
\end{figure}

Figure~\ref{fig_rev_ring} compares the derived  ring-fit radius $R_{\rm torus} \sim  0.16$~pc of NGC~3783 (red)  with eight  interferometric $K$-band radii (blue) reported by \citet{kis09a,kis11a}. These radii are plotted against the UV luminosity $L$, defined as a scaled $V$-band luminosity of $6\ \nu f_{\nu} (V)$, with the $V$ flux  extrapolated from the flux at 1.2 $\mu$m  \citep{kis07}. We can compare these torus radii  with reverberation radii $R_{\rm \tau_K}$ (black) derived from the light traveling distances corresponding to the time lag between the $K$-band  and the UV/optical \citep{sug06}.  They are known to be  proportional to $\sim$$ L^{1/2}$ and are likely probing the dust sublimation radius. The dotted line is the fit curve of the  reverberation radii  (different luminosity values are obtained for the same object  because of  variability and uncertainties of the luminosity derivation). The  reverberation radius of NGC~3783 is $\sim$0.071~pc, which is smaller than the interferometric torus radius $R_{\rm torus} \sim  0.16$~pc (Sect. 3).  Figure~\ref{fig_rev_ring} shows that several  interferometric torus radii  are  larger than the reverberation radii.   Our interpretation is that the interferometric torus radii are averages over the radial dust distribution that emits the $K$-band light, whereas  the reverberation radii probably trace the dust closer to the inner dust torus boundary radius (\citealt{kis09a}).  Furthermore, we  note that $R_{\rm torus} \sim  0.16$~pc is a fit radius calculated with a thin-ring model (i.e., ring width = outer radius minus inner radius = 0). If a dust distribution with a certain thin-ring fit radius  is ring-like and has a ring width larger than zero, then the inner ring radius would be smaller than the thin-ring fit radius.  We have not fitted a ring model with a larger ring width, since the ring width cannot be constrained with the available visibilities. 

\begin{figure}         
\centering \includegraphics[width=6.0cm]{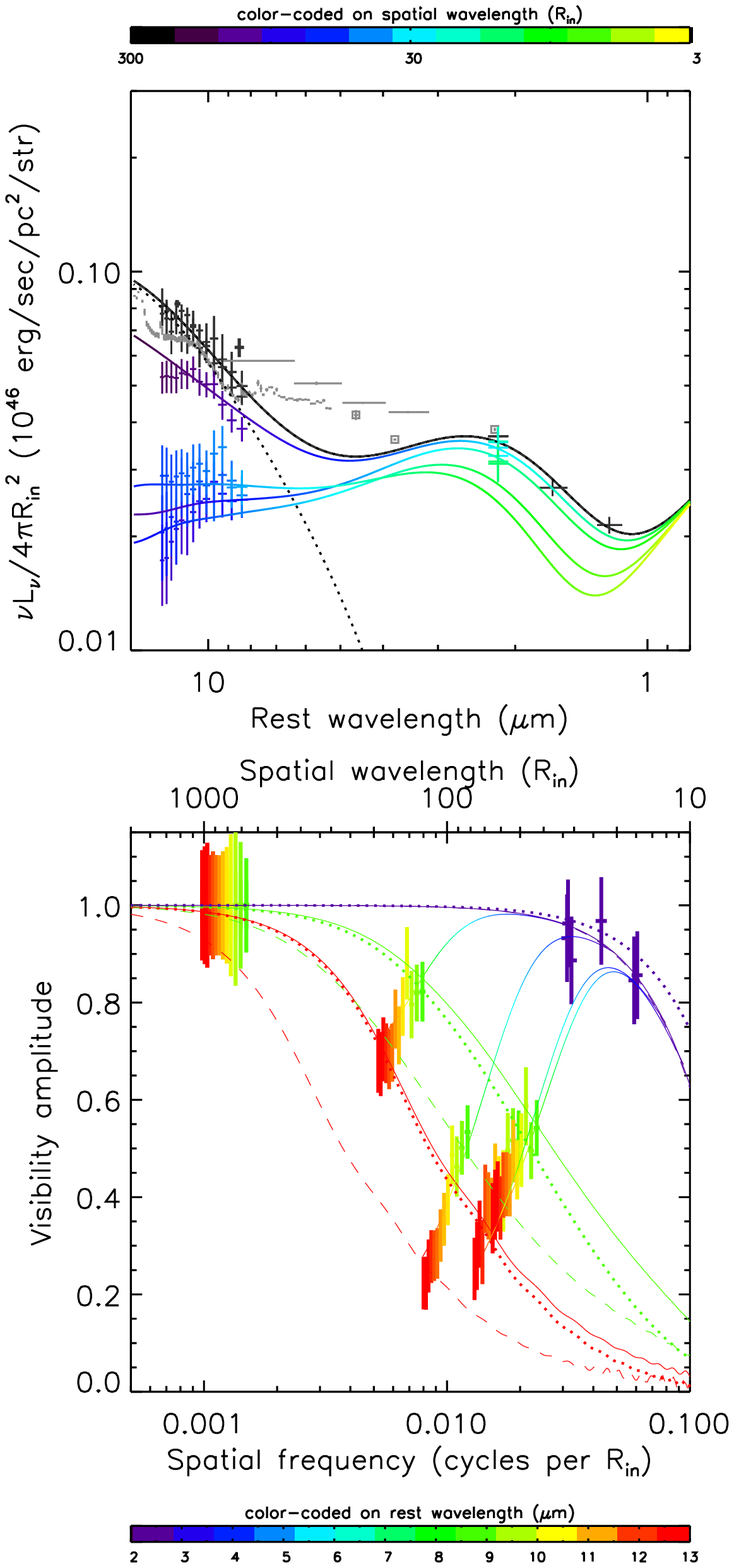} 
 \par\vspace{-0pt}
 \caption{Temperature/density-gradient model including an additional inner hot 1400~K ring component.
{\it Top:} SED observations (black and gray symbols), model SEDs   (black solid line: model including the 1400~K ring component;  black dotted line: model without the 1400~K component),  and  correlated fluxes (yellow, green, and  blue; see \citetalias{kis11b} for more details). The different colors (see top color bar) correspond to different spatial wavelengths measured in units of  the dust sublimation radius $R_{\rm in}$.
{\it Bottom:} New NIR visibilities (purple symbols) and our MIR visibilities from \citetalias{kis11b} (the symbols with colors from green to red correspond to 8.5 to 13~$\mu$m; see color-coding bar; note that the spatial frequency is given in units of cycles per  $R_{\rm in}$). The solid and dashed lines are the visibilities of the temperature/density-gradient model including an additional inner 1400~K ring. The red, green, and purple lines are the model visibilities at 13, 8.5, and 2.2~$\mu$m, respectively  (solid lines: model curves for the PA along the equatorial axis; dashed line: along the polar direction; see \citetalias{kis11b}).  The dotted lines are the visibilities and SED of the same temperature/density-gradient model, but without an inner hot 1400~K ring component. 
}    
\label{fig_model}
\end{figure}
 
\subsection{Simultaneous modeling of the NIR AMBER visibilities, the MIR MIDI visibilities, and the SED}
Mid-infrared (MIR) MIDI interferometry  of NGC~3783 was reported by \citet{bec08}, \citet{kis09b}, and \citet{kis11b} (= \citetalias{kis11b}). For the interpretation of these observations, we  used a temperature/density-gradient model including an additional hot inner ring component with a temperature of 1400~K (\citetalias{kis11b}).  This simple model assumes that the face-on surface brightness distribution of the torus is dominated by the IR radiation from dust clouds directly illuminated and heated by the AD. These dust clouds are probably located near the torus surface since clouds deep inside the torus are not directly illuminated.  

The surface brightness distribution of this temperature/density-gradient model  depends on two distributions, namely a radial temperature and a radial surface density distribution (see Eq. 8 in \citetalias{kis11b}). The maximum dust temperature $T_{\rm max}(r)$ at distance $r$ is assumed to be proportional to $(r/R_{\rm in})^{\beta}$, where $r$ is the radial distance,  ${\beta}$ is the power-law index, and $R_{\rm in}$ is the dust sublimation radius empirically given by  the NIR reverberation radius $R_{\rm \tau_K}$ \citep{gla92}, i.e., we define $R_{\rm in} = R_{\rm \tau_K}$ (\citetalias{kis11b}, Eq. 1). 

Furthermore, the surface brightness distribution depends on the surface density function $f_{\epsilon} (r) = f_0 (r/R_{\rm in})^{\alpha}$ of the heated dust clouds near the surface (power law with index $\alpha$). The emissivity factor $f_{0}$ is equal to $f_{\epsilon} (r)$ at $r$ equal to the  sublimation radius $R_{\rm in}$. Our IR  observations are only  sensitive to the dust  clouds near the surface, which have the temperature $T_{\rm max}(r)$, and not to the cold dust inside the torus. $f_{\epsilon} (r)$ can be regarded as a surface filling factor multiplied by the emissivity (\citetalias{kis11b}). If the emissivity of optically thick illuminated clouds does not depend sensitively on the radial distance from the illuminating source or the observing wavelength  (e.g., see Fig. 3 of \citealt{hoe10}), the factor $f_{\epsilon} (r)$ is roughly proportional to the radial surface density distribution of the heated dust.

Interestingly, the application of this temperature/density-gradient model to several AGNs in  \citetalias{kis11b} and  to the NGC~3783 observations reported in this paper (see  Fig.~\ref{fig_model}) has shown that an additional inner  hot model component is required  with a temperature of 1400~K and a radius of one or a few dust sublimation radii in order to explain all observations. This hot component might play a similar role as the puffed-up inner rim  discovered in several young stellar objects near the dust sublimation radius. 

We simultaneously fitted this temperature/density-gradient model including an inner  1400~K  ring   to  our new $K$-band data as well as  the  MIR data and the SED from \citetalias{kis11b}. The goal of this modeling  is to further constrain physical parameters of the dust distribution. Figure~\ref{fig_model} shows that this model is able to simultaneously reproduce all NIR and MIR visibilities as well as the SED. In Fig.~\ref{fig_model} (bottom), the NIR visibilities (purple) and the MIR visibilities are shown (from green to red, wavelengths 8.5 to 13~$\mu$m; see color-coding bar).  The purple, red, and green curves  are the model visibilities at 2.2, 13, and 8.5~$\mu$m, respectively. The solid lines are the model curves along the PA of the equatorial axis, the dashed line along the polar direction, as defined by optical polarization measurements (see  \citetalias{kis11b}, where this elliptical and a circular symmetric model are presented).

If there is no hot inner ring component added to the above temperature/density-gradient model, then the $K$-band model visibilities (blue dotted line in Fig.~\ref{fig_model}, bottom) are systematically higher than observed and the model SED (black dotted line in Fig.~\ref{fig_model}, top) has a deficiency in the NIR. Therefore, the above inner 1400~K ring component is required to explain the data.   

This new modeling including $K$-band visibilities (Fig.~\ref{fig_model})  is more detailed than that in  \citetalias{kis11b}.  Some of the parameters are similar as in   \citetalias{kis11b}; a temperature power-law   index $\beta = -0.37$, density index $\alpha$  = 0.07, and emissivity factor $f_0 $  = 0.12 for the power-law component of the elliptical model with an  inner temperature of 700~K (\citetalias{kis11b}, Table 8). However, in our new modeling, the emissivity factor  of the inner 1400~K ring is $0.038 \pm 0.016$ and the radius of the  hot  1400~K ring is $2.29 \pm 0.47 R_{\rm \tau_K}$ or $\sim$0.16\,pc   (with the above $R_{\rm \tau_K} = 0.071$~pc), which is no longer fixed to 1~$R_{\rm \tau_K}$ as in \citetalias{kis11b}. This 1400~K ring radius of  $\sim 2.29 R_{\rm \tau_K}$, which is relatively large compared to the reverberation radius $R_{\rm \tau_K}$, is a representative thin-ring radius and not the inner radius of an extended ring (see discussion in Sect. 4.1). This large radius probably indicates a relatively shallow, extended innermost dusty structure  (\citetalias{kis11b}) in NGC~3783.

\section{Conclusion}

We have derived a  torus radius of $0.74 \pm 0.23$~mas or $0.16 \pm 0.05$~pc  (thin-ring fit). To derive this  NGC~3783 torus radius, we  took  into account  an estimated  relative flux contribution of 0.5\% from the underlying host galaxy  in the 60~mas AMBER FOV and 21\% from the unresolved accretion disk. This torus radius is approximately 2.3 times larger than the  $K$-band reverberation radius $R_{\rm \tau_K}$~$\sim$ 0.071~pc (see discussion  in Sect. 4.1).  For the interpretation of the observations, we employed a temperature/density-gradient model including a hot inner 1400~K ring. We simultaneously fitted  our new NIR AMBER visibilities, the MIR MIDI visibilities from \citetalias{kis11b}, and the SED to constrain  physical parameters of the dust distribution.  For the power-law component of the model, we derived a    temperature power-law index $\beta \sim   -0.37$ and a surface density index $\alpha \sim$  0.07. For the   required 1400\,K  ring component, a  radius of $\sim$2.3 reverberation radii or $\sim$0.16\,pc  was found, whereas in the modeling in  \citetalias{kis11b}, the 1400~K ring radius was assumed to be one reverberation radius.   This 1400~K ring radius of $\sim$0.16\,pc, which is relatively large compared to the reverberation radius, is a representative thin-ring radius and not the inner radius of an extended ring. This radius probably indicates a relatively shallow, extended inner dusty structure.  Our study of NGC~3783 and the results in \citetalias{kis11b} show that the simultaneous modeling of both NIR and MIR interferometric observations is a powerful tool for future detailed studies of AGN tori.
 
\begin{acknowledgements}
We thank the ESO VLTI team on Paranal for the excellent collaboration  and  the referee for his valuable comments.  This publication makes use of the SIMBAD database operated at CDS, Strasbourg, France.  
Some of the data presented here were reduced using the publicly available
data-reduction software package {\it amdlib} kindly provided by the Jean-Marie Mariotti Center (http://www.jmmc.fr/data\_processing\_amber.htm).
\end{acknowledgements}

\bibliographystyle{aa}
\bibliography{n3783}
\end{document}